\documentstyle[12pt]{article}
\def\lsim{\mathrel{\rlap{\lower4pt\hbox{\hskip1pt$\sim$}}
    \raise1pt\hbox{$<$}}}                
\def\gsim{\mathrel{\rlap{\lower4pt\hbox{\hskip1pt$\sim$}}
    \raise1pt\hbox{$>$}}}                

\def\beq{\begin{equation}}
\def\eeq#1{\label{#1}\end{equation}}
\def\barr#1{\begin{equation}\begin{array}{#1}\displaystyle}
\def\earr#1{\end{array}\label{#1}\end{equation}}
\def\bit{\begin{itemize}}
\def\eit{\end{itemize}}
\def\ben{\begin{enumerate}}
\def\een{\end{enumerate}}
\def\bce{\begin{center}}
\def\ece{\end{center}}
\def\bmi{\begin{minipage}}
\def\emi{\end{minipage}}
\def\btab{\begin{tabular}}
\def\etab{\end{tabular}}

\def\avv#1{<\!\!#1\!\!>}
\def\r#1{(\ref{#1})}
\def\od#1#2{\displaystyle \frac{#1}{#2}}
\def\lil#1{\hbox to\hsize{ #1 \hfil}}
\def\lilr#1#2{\hbox to\hsize{#1 \hfil #2 }}
\def\lir#1{\hbox to\hsize{\hfil #1 }}
\def\lic#1{\hbox to\hsize{\hfil #1 \hfil}}

\def\h#1{\hspace{#1cm}}

\def\si{\sigma}
\def\ga{\gamma}
\def\ep{\epsilon}
\def\al{\alpha}

\def\de{\delta}
\def\De{\Delta}

\def\Z{\indent\indent}

\begin{document}

\lic{\Large \bf The Systematical Uncertainties in Measurements of}
\vspace{.2cm}
\lic{\Large \bf the Spin-Dependent Structure Function $g_1^n$ with $^3$He}
\vspace{.2cm}
\lic{\Large \bf Target due to Radiative Correction Procedure}

\vspace{.5cm}
\lic{\large I.Akushevich$^1$, A.Nagaitsev$^2$}
\vspace{0.2cm}
\lil{\h5 $^1$ \it NC PHEP, Minsk Belarus}
\lil{\h5 $^2$ \it JINR, Dubna Russia}
\vspace{0.7cm}

\begin {abstract}
The sources of the systematical uncertainties
due to radiative correction procedure
in measurements of the structure function $g_1^n$ with $^3$He target
are considered.
Their numerical estimations
are presented. The relative systematical uncertainty
does not exceed 5\%.

\end {abstract}

\input{epsf}

\Z

The measurements of the spin-dependent structure functions performed
in the last years (SMC, E154/155)
have the tendency to decrease essentially  the
statistical uncertainties comparing with previous ones (EMC, E142/143).
Thus the one of main problem is to decrease the magnitude of
the systematical uncertainties or define more correct way of
their calculation. In general systematical uncertainties come
from
measured quantities, namely: values of the beam and target polarizations,
efficiencies of the coordinate detectors etc. There are also some
contributions in total systematic uncertainty due to unmeasured
quantities such as radiative corrections (RC) to be applied
to extract the one-photon exchange cross section from the measured
one.

So called radiative events,
which
originate from loop diagrams and
from processes with the emission of additional real photons,
 cannot completely be removed by experimental methods and so
they have to be calculated theoretically and subtracted
from measured cross sections.
The calculation of the radiative corrections requires knowledge
of spin-independent and spin-dependent sturcture functions both in region
measured in
the considered experiment and beyond it.
The choice of  different parameterizations of the sturcture functions,
elastic and quasielastic formfactors, neglecting of electroweak and
higher oreder effects and simplifications in RC procedure
leads to
uncertainties in  calculations of RC.
 The approach of
calculation of these systematical uncertainties is presented in
this report.  We consider the case of the $^3$He target and
kinematical region close to HERMES
\cite{HERMES} and SLAC \cite{slac} experiments.
To get the numerical estimations of the systematical uncertainties
due to radiative correction procedure we use the special program
\cite{AkNa} and radiative correction calculation code
POLRAD 2.0 \cite{POLRAD20,ASh}.

The extraction of the spin asymmetry $A_1(x,Q^2)$ in
the experiments
measuring spin-dependent structure functions is based on the
following formula
\beq
A_1(x,Q^2)=\od1{\avv{Df_d}}{
N^{\uparrow\downarrow}-N^{\uparrow\uparrow}
\over
N^{\uparrow\downarrow}+N^{\uparrow\uparrow}
},
\eeq{001}
where $\avv{Df_d}$ is mean value of product of depolarization
and dilution factors (see \cite{DYM} for details),
$N^{\uparrow\uparrow}$ and $N^{\uparrow\downarrow}$
are
  the number of events for
parallel
and antiparallel spin target
configurations.

In order to take into account the radiative effects
these numbers have to be calculated as the weighted sum
\beq
N^{\uparrow\downarrow,\uparrow\uparrow}=
\sum_{\uparrow\downarrow,\uparrow\uparrow}w(x,Q^2).
\eeq{0015}
Here
the weight ($w$)  is calculated as a ratio of Born
and observed cross sections
\beq
w=\od{\si_{0}}{\si_0+\si_{RC}}
    \eeq{002}
We will refer to this
procedure as 'exact'. 

The exact procedure requires the RC
calculation for each event, so in practice other scheme is used.
We will refer to this procedure as 'standard'. In this case
the RC is applied to
asymmetry averaged  at $x$ bins.

The radiative correction $\De A_1$ to the measured
asymmetry is defined as:
\beq
A_{1\;meas}=A_{1}+\De A_1
\eeq{rc01}
and can be written in terms of spin-independent($\si^{u}$)
 and spin-dependent ($\si^{p}$) parts
of  DIS cross section. 
\beq
\De A_1=
{\si_0^u\bigl(\si^p_{in}(g_1)+\si^p_{q}+\si^p_{el}\bigr)
-\si_0^p(g_1)\bigl(\si^u_{in}+\si^u_{q}+\si^u_{el}\bigr)
\over
\si_0^u\bigl((1+\de_v)\si_0^u+\si^u_{in}+\si^u_{q}+\si^u_{el}\bigr)
},
\eeq{rc02}
 where $\de_v=\si_v^p/\si_0^p=\si_v^u/\si_0^u$.
 The polarized parts of Born cross section and
inelastic radiative tail depend
on $g_1$.

The radiative correction procedure is
performed as follows.

1. The measured asymmetry $A^{m}_{1\; i}$ ($i=1,...,N_x$, where
$N_x$ is number of $x$ bins)
is fitted by a function\footnote{The example such a function
 for $^3$He target can be found  in Appendix of ref \cite{DYM}}
 with taking into account statistical uncertainties of $A^{m}_{1\; i}$.

2. The constructed fit is used for calculation of $\si^p_{in}(g_1)$
and $\si_0^p(g_1)$.

3. The extracted asymmetry is calculated for each kinematical bin as
follows
\beq
A_{1\; i}^{ext}=A^{m}_{1\; i}
-\De A_1
\eeq{03}

There are three important sources for uncertainties coming from
radiative correction procedure: a) using of simplified (standard) scheme
instead of the exact one ; b) using of models and data for
structure functions; c) physical effects which are neglected in
the standard scheme.

\begin{figure}[htb]
\unitlength 1mm
\begin{picture}(160,160)
\put(0,15){
\epsfxsize=16cm
\epsfysize=16cm
\epsfbox{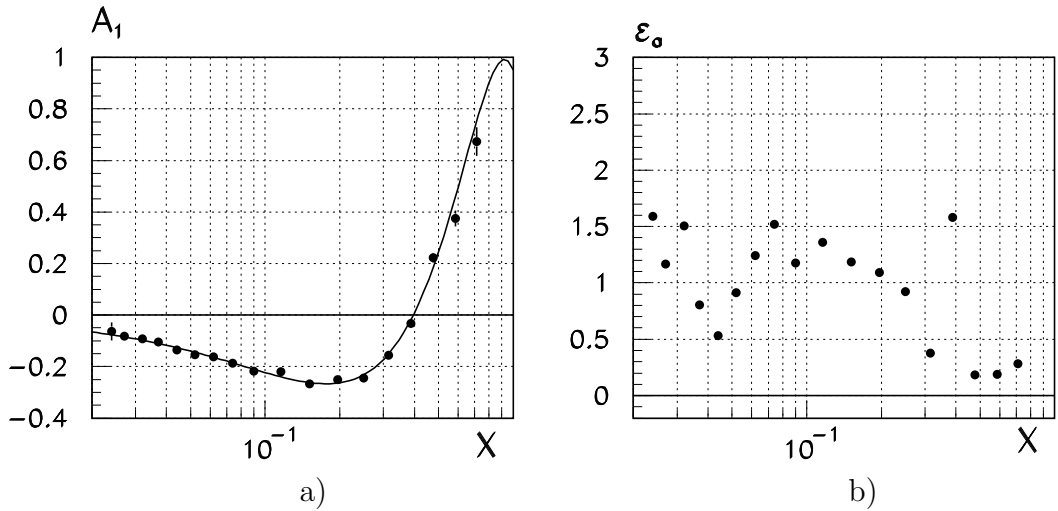}
}
\put(45,92){{\rm a)}}
\put(118,92){{\rm b)}}
\end{picture}
\vspace{-9.5cm}
\caption{\protect\it
 The generated asymmetry along used fit and
relative systematical uncertainties versus $x$
 in percents due to simplification procedure. 
  It should be noted that peak for $x\protect\sim 0.4$ 
is artificial and originates from the fact
that neutron spin asymmetry is close to  zero in this
kinematical point.
}
\label{f1}
\end{figure}

To calculate the systematical uncertainties  of the types a),
b) and c)
the following scheme is used.
The Monte-Carlo kinematical
events are generated according to random flat generator:  $\ln Q^2$ for
$Q^2$ over allowed $Q^2$-- region and flat for
$\nu$ over allowed $\nu$ -- region.
 After calculation of kinematic variables and applying of kinematic
cuts close to acceptance of the HERMES experiment \cite{HERMES},
the weights with DIS cross section
$w^{1\ga}$
are obtained at given kinematical point.
The model for spin asymmetry  and generated
Born asymmetry are plotted on fig.\ref{f1}a.

For each event the
recalculation of the weight is performed with radiative correction factor:
\beq
w^{obs}=\od{w^{1\ga}}{w},
\eeq{0003}
where weight $w$ defined from eq. \r{002} is
calculated by code POLRAD 2.0.
The measured ($A_1^{obs}$) and extracted within exact scheme
($A_1^{ex}$) asymmetries are obtained using formulas
\r{001} and \r{0015} with the weights $w^{obs}$ and $w^{1\gamma}$.

 The
application of the standard scheme to $A_1^{obs}$ gives the
extracted asymmetry $A_1^{st}$
within standard scheme.
To estimate the relative
 systematical uncertainty of type a), we calculate
the difference
between extracted asymmetries within the exact ($A_1^{ex}$) and
standard ($A_1^{st}$) schemes for each kinematical bin:
\beq
\ep_a={\left| A_1^{ex}-A_1^{st}\right| \over A_1^{ex}}.
\eeq{0047}
The relative systematical
uncertainties of types b) and c)  are
estimated as follows
\beq
\ep_{b,c}={\left| A_1^{ex}-{\tilde A}_1^{ex}\right| \over
A_1^{ex}}.
\eeq{0048}
The asymmetry ${\tilde A}_1^{ex}$ is calculated using the weight
${\tilde w}^{ex}=w^{obs}{\tilde w}$. The weight ${\tilde w}$
 is also calculated
 according to formula \r{002}, but $\si^{obs}$
 is calculated
with different models for structure functions or with taking into account
electroweak and high order effects.

The quantity $\epsilon_a$ can be non-zero only due to order of
averaging within standard and exact schemes for the same models
for structure functions.
In case of 'standard' scheme, the results are calculated for
averaged kinematical point at each bin. But this mean value
can be shifted due to  very different dependence of Born and radiative
corrected cross sections on kinematical variables.
The exact scheme is free from this shortcoming.
In this case the RC
is taken into account before averaging using
calculation of the weight \r{002} for each event.
The relative systematical shift obtained with
standard and exact schemes
does not exceed 2\% (fig.\ref{f1}b).

The quantity $\si_0^u$ in eq.\r{rc02} depends on $F_2$ in
the given kinematical point
$Q^2$ and $x$, but $\si_{in}^u$ (see eq. \r{rc02}) requires
knowledge of the structure function $F_2$ in the wide region of
varying of kinematical variables $x$ and $Q^2$. 
So the
fit of $F_2$ used for RC calculation has to  describe adequately
both resonance and
DIS region as well as to have correct asymptotics behaviour for
$Q^2\rightarrow
0$ and
$W^2\rightarrow (M+m_{\pi})^2$. Such  fit can be
constructed
on the basis of NMC parameterization of $F_2$ for protons and
deuterons
\beq
F_2^{^3 \rm He}(x,Q^2)=\od13 \biggl(F_2^{H}(x,Q^2)+F_2^{D}(x,Q^2)
.
\biggr) \eeq{0028}
$F_2^p$ and $F_2^d$ are taken from fit described in \cite{NMC}. The
fit takes into account the contribution of $\De$ resonance, has
correct behaviour on boarders and describes DIS data for $x\gsim
0.01$.

Another possibility is to change the model for
$F_2$ on standard POLRAD fit \cite{ASh}, which includes Brasse
parameterization of the three resonances (instead of one resonance
as in standard fit \r{0028}), Stein fit for small $Q^2$ region and
15 parameters
NMC fit for DIS region. 
The final results 
shows that
unpolarized structure function can change asymmetry by approximately
$1\%$.

Both polarization and unpolarization contribution of quasielastic
radiative tails  depend on quasielastic
response functions ${\cal F}_i^{q}$ (see \cite{POLRAD20} for details).
The electric and magnetic formfactors for
proton and neutron fall as $Q^8$
for high
$Q^2$. So the only region of small $Q^2 \sim M^2$ is
important.
In this region the nucleon
formfactors are known with good accuracy, and their variation does
not lead to systematical uncertainty. To the contrary the
models for
electric $S_E$, magnetic $S_M$  and mixed $S_{EM}$ suppression factors 
 differ in this region. The code POLRAD 2.0 exploits
Y-scaling
hypothesis \cite{West,Tho} $S_M=S_E=S_{EM}=F(\nu_0)$ and scaling
function $F(\nu_0)$ are calculated
 in Fermi gas model \cite{Wal,Mon}.
Alternatively the suppression factors can be also calculated
within the  sum  rule  approach  \cite{LLS}.
The difference is important for small values
of $x$, which correspond to
high $y$, where it can reach 1.5\%.

The elastic structure functions are calculated as
quadratic combinations of electric and magnetic formfactors.
A simple Schiff's model with
gaussian wave function \cite{Shi} is used as an alternative model.
The results are similar to quasielastic case.

Resonance region gives a large contribution to RC for
spin-independent DIS.
The contribution to resonance region in $g_1$ can be also
important.
 Unfortunately, there are no enough experimental data  or
satisfactory models  for $g_1^n(x,Q^2)$ in resonance region
($W^2<4$).
 It is the main
reason why scaling behaviour of spin asymmetry is extrapolated
into
resonance region under POLRAD consideration. For alternative
approach
 we
used
simple model for structure function $g_1^n(x)$ in region of $\De$(1232)
resonance
constructed on the basis of two assumptions, namely:
the  $W^2$-dependence
has the Gaussian form with height and width which can be
roughly estimated from recent SLAC data \cite{SLACresnew},
the $Q^2$-dependence is
defined by resonance contribution to
Drell-Hearn-Gerasimov sum rule given in ref.\cite{Lider}.
The  relative systematical uncertainty due to resonance region is
important ($\sim$1.5\%) for small $x$-region.

The kinematical coefficient front of structure function $g_2$ is small
enough
for both at
the Born level and for RC cross section. So, the
contribution of the structure function is neglected normally. To study
influence
of such an approach the model of
the Wandzura-Wilczek is applied \cite{WW}. 
The small effect is obtained. The exception is last two bins where
$\epsilon_b \sim 1$\%.

The electroweak effects are not included in standard
radiative correction procedure, because for
 current polarized experiments
 $Q^2 \sim 10$GeV$^2 \ll M_z^2$ ($M_z$ is the $Z$-boson mass) and
hence their contribution
is small, but it has to be added to systematical error.
Such a systematical uncertainty is estimated
at the born level
using
 code POLRAD 2.0.
The electroweak
correction cannot be calculated by model independent way, that is
why the quark parton model was used and  GRV- ,
GRSV-parameterizations
\cite{GRV,GRSV96} for spin-independent and spin-dependent partonic
distributions were applied.
The correction is important ($\epsilon_c \sim 1.5\%$)
for high $x$. 

In standard consideration  the effects of higher order is
estimated by simple exponentiation procedure of soft photons
\cite{POLRAD20,Sh}. The
POLRAD 2.0 gives also a possibility to obtain the $\al^2$ order
correction within structure function approach \cite{Kuraev,Spies}.
The radiation is considered to be collinear. There are three
possibilities: initial, final state radiation and contribution of
the Compton process. The first and second are important for
inelastic radiative tail. The last one is extremely important for
elastic and quasielastic radiative tails. 
The  relative systematical uncertainties due to the higher order effect
can exceed
2\% at low $x$-region.

 The systematical uncertainties considered in previous sections
can be gathered together to obtain the total relative  contribution
due to RC to systematical uncertainties.
The following systematical uncertainties are considered to be independent
\bit
\item simplification of procedure
\vspace{-0.3cm}\item unpolarized structure function $F_2$ 
\vspace{-0.3cm}\item quasielastic structure function 
\vspace{-0.3cm}\item elastic formfactors 
\vspace{-0.3cm}\item resonance region of structure function $g_1$
\vspace{-0.3cm}\item polarized structure function $g_2$ 
\vspace{-0.3cm}\item electroweak effects 
\vspace{-0.3cm}\item higher order effects 
\eit

\begin{figure}[thb]
\unitlength 1mm
\begin{picture}(160,160)
\put(0,15){
\epsfxsize=16cm
\epsfysize=16cm
\epsfbox{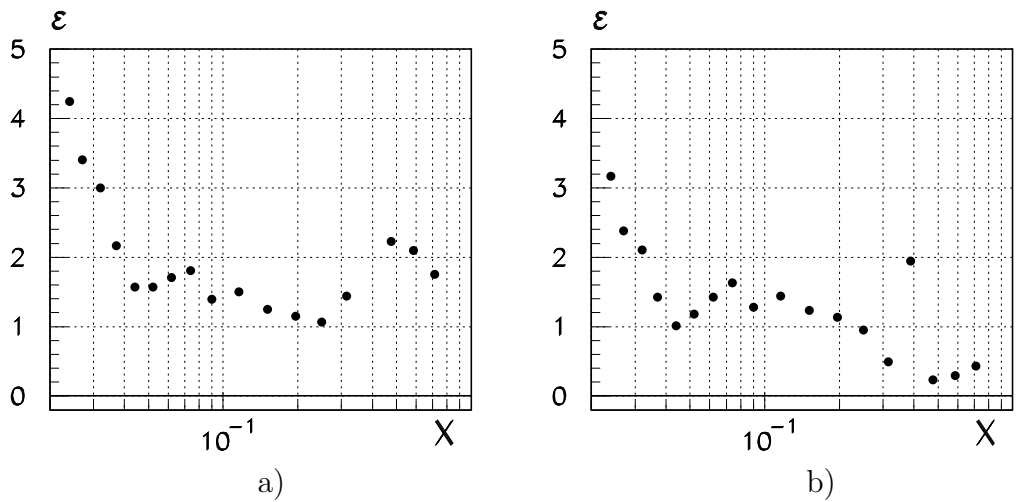}
}
\put(45,92){{\rm a)}}
\put(118,92){{\rm b)}}
\end{picture}
\vspace{-9.5cm}
\caption{\protect\it
The relative systematical uncertainties due to radiative corrections.
}
\label{f5}

\end{figure}
\vspace{1cm}

The quadratic sum of the uncertainties mentioned above
presents
in fig.\ref{f5}a. For $x\gsim 0.3$ the total systematical
error does not exceed 2\%. However for the first $x$ bins with $x\lsim
0.3 $ and $y\sim 0.85$ the effect is larger and can reach 5\%.

Note that the uncertainties of types c) can be rejected from total
sum, if they are included in standard scheme. For  higher order
correction it can be done without additional assumption, but
electroweak correction calculation requires  usage of the
quark-parton model.
In this case systematical uncertainties  due to RC come only from
procedure simplification and uncertainties in structure functions. This
result
is presented on fig.\ref{f5}b. It does not exceed
4\%.

{\bf Acknowledgements.}
We are grateful to N.Shumeiko and  I.Savin for help and support. Also we
would
like to thank N.Akopov, N.Gagunashvili, V.Krivokhigine, P.Kuzhir and
D.Ryckbosch
for useful discussion and comments.


\end{document}